# Complete 72-Parametric Classification of New and Old Kinds of Surface Plasmon Waves

## MAXIM DURACH[1],*

[1]*Department of Physics and Astronomy, Georgia Southern University, 65 Georgia Avenue, Statesboro, GA 30461*
*Corresponding author: mdurach@georgiasouthern.edu*

**We propose a general and complete classification of all possible new and old kinds of surface plasmon waves that can propagate at boundaries of arbitrary linear, local bi-anisotropic media, including the quartic metamaterials. For arbitrary frequency, wavelength, propagation direction, penetration depths and fields of the proposed surface plasmon waves we found the dispersion condition and determined the 72-parametric class of media that support a particular surface plasmon. A member of each class is a pair of anisotropic materials without magnetoelectric couplings.**

In this letter we introduce a new broad class of surface electromagnetic waves, which propagate at interfaces of arbitrary bi-anisotropic media with magnetoelectric coupling, which includes the quartic metamaterials [1]. Surface electromagnetic waves have been studied since the time of Sommerfeld [2] and Zenneck [3]. Zenneck's solution of Maxwell's equations is based on matching of evanescent waves at a planar boundary of two media and was originally applied to the interface between earth and air. Later it was realized by Fano [4], Ritchie [5] and Stern [6] that Zenneck's solution describes the surface plasmon polaritons at metal-dielectric interfaces. Generally speaking, Zenneck waves correspond to a broad family of surface polariton waves at boundaries of isotropic media [7]. With the advent of metamaterials and photonic crystals the design of custom anisotropic media became possible and more complex surface electromagnetic waves became of interest, such as D'yakonov waves [8-12], optical Tamm waves [13-19], Dyakonov-Tamm modes [20-22], and hyperbolic Tamm plasmons [23,24].

The attention to new metamaterials is driven by the possibilities of novel optical effects, such as negative refraction, hyperbolic dispersion, optical magnetism non-reciprocity etc [25,26], and is following the trajectory of increasing complexity of the effective media that describe these metamaterials [27]. Nevertheless in most cases the effective media under consideration are described by at most 9 parameters [11,28,29]. Recently we made the next logical step in this development and considered the most general linear, local bi-anisotropic metamaterials with k-surfaces described by 4th order polynomials in k-vector components, i.e. quartic metamaterials [1]. We have shown that quartic media support plane waves with arbitrary polarization and wavelength and that the 36 effective medium parameters can be retrieved from the amplitudes and k-vectors of 6 plane waves propagating in the medium [1]. In this letter we predict that a boundary between two quartic metamaterials supports a new kind of surface electromagnetic waves, if the condition we provide below [Eq. (7)] is met [see Fig. 1].

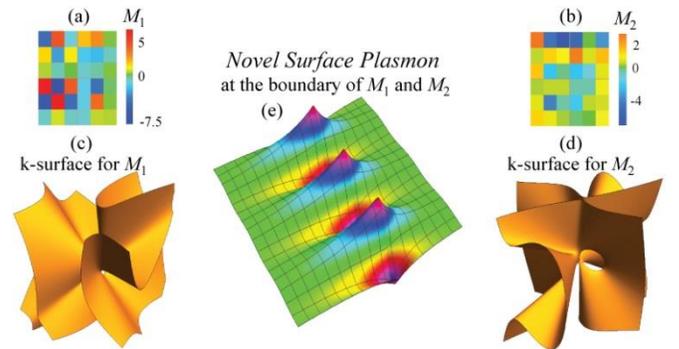

Fig. 1. New kind of surface plasmons propagating between a pair of quartic media $M_1$ and $M_2$. (a-b) The constitutive effective parameter matrices for the media. (c-d) The corresponding k-surfaces with real $k_x, k_y, k_z$. (e) The longitudinal electric field distribution for the proposed surface plasmon wave propagating between $M_1$ and $M_2$.

The constitutive relationships between $D, B$ and $E, H$ are represented by a 6x6 transformation matrix $\widehat{M}$ (i.e. 36 parameters) and the quartic metamaterials correspond to the most general linear local constitutive relationship [30-35,1]

$$\begin{pmatrix}D\\B\end{pmatrix} = \widehat{M}\begin{pmatrix}E\\H\end{pmatrix}, \widehat{M} = \begin{pmatrix}\hat{\epsilon} & \hat{X}\\\hat{Y} & \hat{\mu}\end{pmatrix}, \qquad (1)$$

where $\hat{\epsilon}, \hat{\mu}, \hat{X}, \hat{Y}$ are 3x3 tensors characterizing dielectric permittivity, magnetic permeability and magnetoelectric coupling correspondingly. The examples of matrices $\widehat{M}$ are color-coded in Fig. 1(a-b) for quartic media $M_1$ and $M_2$. Note, that here we do not discuss the immediate availability with the current technology of the values of material parameters we obtain in our examples, but provide a recipe to obtain the values of the parameters needed for the desired surface plasmon waves to inform and drive the future design of the corresponding metamaterials.

Plane waves in quartic media follow Maxwell's equations

$$\boldsymbol{k} \times \boldsymbol{E} = k_0 \boldsymbol{B} \text{ and } \boldsymbol{k} \times \boldsymbol{H} = -k_0 \boldsymbol{D}, \qquad (2)$$

where vectors $\mathbf{D}, \mathbf{B}$ and $\mathbf{E}, \mathbf{H}$ follow the constitutive relations (1). The Eqs. (1)-(2) can be rewritten as [1]

$$\hat{Q}\Gamma = \hat{M}\Gamma, \quad (3)$$

with $\Gamma = (E_x, E_y, E_z, H_x, H_y, H_z)$ and

$$\hat{Q} = \begin{pmatrix} 0 & -\hat{R} \\ \hat{R} & 0 \end{pmatrix}, \hat{R} = \frac{1}{k_0}\begin{pmatrix} 0 & -k_z & k_y \\ k_z & 0 & -k_x \\ -k_y & k_x & 0 \end{pmatrix}.$$

This system has nontrivial solutions if the determinant $\det(\hat{M} - \hat{Q})$ is zero which is equivalent to a quartic equation with respect to the components of the k-vector [1]

$$\sum_{i+j+l+m=4}[\alpha_{ijlm}k_x^i k_y^j k_z^l k_0^m] = 0 \quad (4)$$

with 35 coefficients $\alpha_{ijlm}$. In the sum the powers $i, j, l, m$ run from 0 to 4 such that $i + j + l + m = 4$. For real $k_x, k_y, k_z$ the quartic Eq. (4) describes a quartic k-surface, corresponding to solutions $\Gamma$ of Maxwell's Eqs. (2)-(3). The examples of quartic k-surfaces are shown in Fig. 1(c-d) and correspond to metamaterials represented by matrices $M_1$ and $M_2$ color-coded in Fig. 1(a-b).

In Fig.1(e) we show a field distribution example of the proposed surface plasmon wave that propagates between quartic media $M_1$ and $M_2$ in Fig. 1(a-b). Below we outline the algorithm of finding pairs of quartic media that support plasmons with customizable frequencies, wavelengths, propagation direction, penetration depths and fields.

Consider two quartic media, whose interface is in $xy$-plane at $z = 0$. Due to translational symmetry along $x$ and $y$, solutions of Maxwell's equations at this interface can be classified according to the longitudinal k-vector $\mathbf{k}_\parallel = (k_x, k_y)$. For fixed $k_x$ and $k_y$ the dispersion Eq. (4) is a quartic equation of one variable $k_z$. In general, the 4 roots of this equation are complex $\mathbf{k} = (k_x, k_y, k_z' + ik_z'')$ and represent evanescent waves, decaying or growing in the z-direction. If $\alpha_{ijlm}$ are real the complex roots of this equations come in complex conjugated pairs with opposite attenuation as follows from the complex conjugate root theorem [36], one root representing decay along z-axis and another - unbounded growth. The polarizations of the paired waves are related, which can be seen from Eq. (3). For example, if matrix $\hat{M}$ is real $M_{ik}^* = M_{ik}$, then if the fields of one of the waves are given by $\Gamma$, the paired wave should have $\Gamma^*$ for the fields.

Now let us assume that we would like to construct a surface plasmon with custom wavelength propagating in a desired direction, i.e. with arbitrary $\mathbf{k}_\parallel = (k_x, k_y)$. In each bounding medium we can select 2 pairs of evanescent waves with desired penetration depths $l = 1/k_z''$. The other roots of Eq. (4) with $\mathbf{k}_\parallel$, should be the diverging waves, corresponding to the selected evanescent waves. As a result, we can construct a surface wave from the selected evanescent waves, whose k-vectors are

$$\mathbf{k}_i^{(1)} = (\mathbf{k}_\parallel, k_{zi}^{(1)'} + ik_{zi}^{(1)''}), i = 1-2 \text{ for } z > 0 \quad (5a)$$

$$\mathbf{k}_i^{(2)} = (\mathbf{k}_\parallel, k_{zi}^{(2)'} - ik_{zi}^{(2)''}), i = 1-2, \text{for } z < 0 \quad (5b)$$

We also select custom field components of these waves as $\Gamma_i^{(1,2)} = (\Gamma_{\parallel i}^{(1,2)}, \Gamma_{\perp i}^{(1,2)})$, where $\Gamma_{\parallel i}^{(j)} = (E_{xi}^{(j)}, E_{yi}^{(j)}, H_{xi}^{(j)}, H_{yi}^{(j)})$ are longitudinal fields and $\Gamma_{\perp i}^{(j)} = (E_{zi}^{(j)}, H_{zi}^{(j)})$ are transverse fields. The boundary conditions for Maxwell's equations require the continuity of the longitudinal components of the fields, which results in a system of equations:

$$A\,\Gamma_{\parallel 1}^{(1)} + B\Gamma_{\parallel 2}^{(1)} = C\,\Gamma_{\parallel 1}^{(2)} + D\Gamma_{\parallel 2}^{(2)}, \quad (6)$$

where coefficients $A, B, C, D$ are the amplitudes of the evanescent waves composing the surface plasmon. The dispersion equation of this novel surface plasmon wave propagating at the boundary of quartic media corresponds to vanishing of the determinant of the system (6)

$$\begin{vmatrix} E_{x1}^{(1)} & E_{x2}^{(1)} & -E_{x1}^{(2)} & -E_{x2}^{(2)} \\ E_{y1}^{(1)} & E_{y2}^{(1)} & -E_{y1}^{(2)} & -E_{y2}^{(2)} \\ H_{x1}^{(1)} & H_{x2}^{(1)} & -H_{x1}^{(2)} & -H_{x2}^{(2)} \\ H_{y1}^{(1)} & H_{y2}^{(1)} & -H_{y1}^{(2)} & -H_{y2}^{(2)} \end{vmatrix} = 0. \quad (7)$$

For example, at a boundary of two isotropic media with different dielectric permittivities $\varepsilon_1, \varepsilon_2$ and magnetic permeabilities $\mu_1, \mu_2$, the fields of evanescent waves propagating in the x-direction can be written as

$$\Gamma_{\parallel 1}^{(j)} = \left(\pm\frac{ik_{zj}''}{k_0\varepsilon_j}, \frac{k_x}{k_0\varepsilon_j}, 0, 0\right) \text{ for TM polarization,}$$

$$\Gamma_{\parallel 2}^{(j)} = \left(0, 0, \pm\frac{ik_{zj}''}{k_0\mu_j}, -\frac{k_x}{k_0\mu_j}\right) \text{ for TE polarization,}$$

turning Eq. (7) into the well-known equation

$$\left(\frac{k_{z1}''}{\varepsilon_1} + \frac{k_{z2}''}{\varepsilon_2}\right)\left(\frac{k_{z1}''}{\mu_1} + \frac{k_{z2}''}{\mu_2}\right) = 0,$$

which gives the dispersion of the conventional surface plasmon polaritons [37]. It is straightforward to obtain the dispersion of Dyakonov plasmons from Eq. (7) as well (cf. Eq. (1) in Ref [10]).

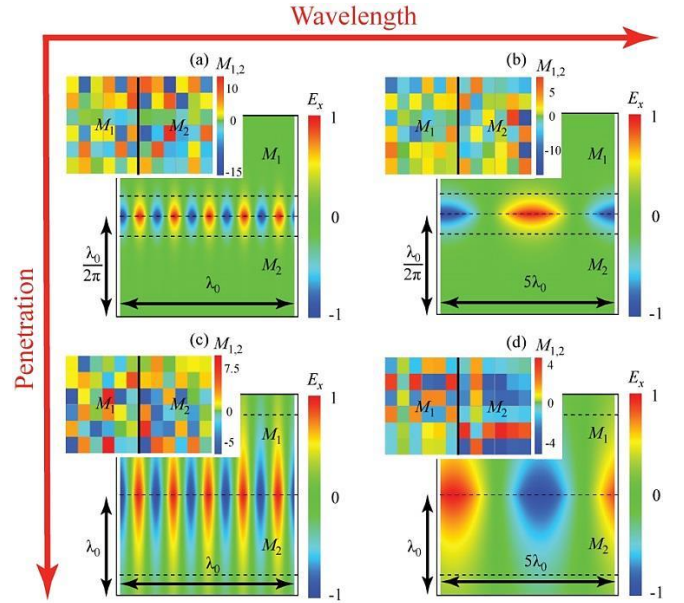

Fig. 2. Examples of the proposed surface plasmons with short and long wavelengths and penetration depths.

After selecting the propagation direction, wavelength, penetration depths and fields of the proposed surface plasmon, with the only limitation given by the dispersion Eq. (7), we can find the effective medium parameters of the quartic metamaterials that support the desired surface plasmon wave at their boundary. To do this we apply the solution of the inverse problem of quartic photonics that we have recently proposed [1]. This routine can be briefly expressed as follows. If one needs to find a material that supports 6 plane waves with custom k-vectors $\mathbf{k}_i$ and polarizations $\Gamma_i$, $i = 1 - 6$, one can construct matrix $\hat{G}$, whose columns are $\Gamma_i$ and matrix $\hat{P}$, whose columns $\Pi_i = (D_x, D_y, D_z, B_x, B_y, B_z)$ are found from Eq. (2). Then the effective parameter matrix $\hat{M}$ of the material is given by [1]

$$\hat{M} = \hat{P}\hat{G}^{-1} \qquad (8)$$

To find the quartic media that support the desired surface plasmon waves we select the values of the k-vectors (5) and their complex conjugates and select the corresponding fields $\Gamma_i^{(1)}, \Gamma_i^{(2)}$ and their complex conjugates. This gives $\Gamma$ and $\Pi$ vectors for 4 waves in each media. The other 2 pairs of waves can be selected freely, as long as they do not have the same $\boldsymbol{k}_\parallel$ as the proposed surface plasmon wave (this is because Eq. (4) is quartic). This freedom defines the *36-parametric class* of different quartic media that all support the selected surface plasmon wave as we discuss in more detail at the end of this letter. After selection of these values Eq. (8) should be used to find the quartic media $M_1$ and $M_2$ that support the desired surface plasmon wave [see Fig. 1].

The examples of our calculations are shown in Figs. 2-3. In Fig. 2 we demonstrate that boundaries supporting surface plasmon modes with arbitrary wavelength and penetration depths can be found using our approach. In Fig. 2(a) the longitudinal electric field $E_x$ of a surface plasmon with short wavelength $\lambda = \lambda_0/5$ and small penetration depths $l = \lambda_0/5$ into both media is plotted. The corresponding materials $M_1$ and $M_2$ are color-coded in the inset of Fig. 2(a) and were found by randomly selecting all of the fields $\Gamma_i$, except for one of the components in the determinant (7), which is adjusted so that (7) is satisfied. Analogously, we found the boundary that supports a surface plasmon with the same penetration, but long wavelength $\lambda = 5\lambda_0$ as shown in Fig. 2(b). Figs. 2(c-d) show the modes and the corresponding boundaries in the insets for the same wavelengths $\lambda = \lambda_0/5$ and $\lambda = 5\lambda_0$ correspondingly, but for a large penetration depth of $l = 5\lambda_0$.

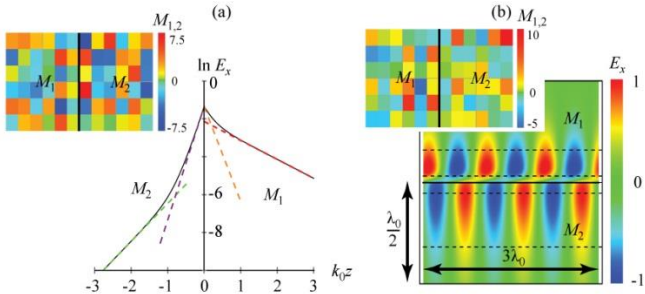

Fig. 3. Interference effects of the evanescent waves composing the proposed surface plasmons. (a) The longitudinal field of a surface plasmon that shows bi-exponentiality. (b) A surface plasmon with the phase shift between the evanescent waves in $M_1$.

Conventional surface plasmons are composed of pairs of evanescent waves, one in each meadium. An important property of the proposed surface plasmons is, generally speaking, the bi-exponentiality in each bounding medium. It is demonstrated in Fig. 3. In panel (a) we show in logarithmic scale the longitudinal field $E_x$ of a surface plasmon propagating at the boundary color coded in the inset. As one can see, in each medium the evanescence is bi-exponential with fast evanescent component overlaying a slowly decaying component. Such behavior in a transverse cross-section can be seen, if the evanescent waves composing the proposed surface plasmon waves are not phase shifted with respect to each other. An example, when they are phase shifted is demonstrated in Fig. 3(b). The evanescent waves in medium $M_2$ are not phase shifted, but there is a phase shift in medium $M_1$.

As was mentioned above, the properties of the proposed surface plasmon wave are determined by the k-vectors and field vectors $\Gamma$ for the pairs of evanescent waves in each medium. If matrix $\hat{M}$ is real this automatically defines the parameters of the 2 waves that are growing away from the boundary into each medium. Nevertheless, to fully define the matrices $\hat{M}$ one needs to choose k-vectors and field vectors $\Gamma$ for 2 extra waves in each medium, i.e. at least 36 parameters, which do not affect the properties of the proposed surface plasmon mode. From this, we can conclude that each proposed surface plasmon mode corresponds to a broad *36-parametric class* of material pairs that support this particular surface plasmon.

We reformulate this as follows. We could require existence of less than 6 waves to define the effective medium parameters matrix $\hat{M}$ in each medium and complement the truncated Eq. (8) with restrictions on material parameters. For example, we require existence of 3 waves in a metamaterial and matrices $\hat{G}$ and $\hat{P}$ in Eq. (8) would have 3 columns instead of 6.

Let us now fix the magnetoelectric couplings $\hat{X}$ and $\hat{Y}$, for both bounding media, i.e. 36 parameters. Then Eq. (8) is rewritten as

$$\hat{\epsilon} = -\hat{P}_H \hat{E}^{-1} - \hat{X}\hat{H}\hat{E}^{-1}, \quad \hat{\mu} = \hat{P}_E \hat{H}^{-1} - \hat{Y}\hat{E}\hat{H}^{-1}. \qquad (9)$$

This means that the permittivity and permeability matrices $\hat{\epsilon}$ and $\hat{\mu}$ can be expressed via the matrices $\hat{X}$ and $\hat{Y}$, and the k-vectors and the field amplitudes of 3 waves in each medium, where

$$k_0 \hat{P}_E = (\boldsymbol{k_1} \times \boldsymbol{E_1}, \boldsymbol{k_2} \times \boldsymbol{E_2}, \boldsymbol{k_3} \times \boldsymbol{E_3})$$
$$k_0 \hat{P}_H = (\boldsymbol{k_1} \times \boldsymbol{H_1}, \boldsymbol{k_2} \times \boldsymbol{H_2}, \boldsymbol{k_3} \times \boldsymbol{H_3})$$
$$\hat{E} = \begin{pmatrix} E_{x1} & E_{x2} & E_{x3} \\ E_{y1} & E_{y2} & E_{y3} \\ E_{z1} & E_{z2} & E_{z3} \end{pmatrix} \text{ and } \hat{H} = \begin{pmatrix} H_{x1} & H_{x2} & H_{x3} \\ H_{y1} & H_{y2} & H_{y3} \\ H_{z1} & H_{z2} & H_{z3} \end{pmatrix}.$$

These 3 waves can be selected to be the 2 evanescent waves composing the proposed surface plasmon plus 1 growing mode corresponding to one of these evanescent waves, which defines the matrices $\hat{E}, \hat{H}, \hat{P}_H, \hat{P}_E$. After this variation of the magnetoelectric couplings $\hat{X}$ and $\hat{Y}$ in Eq. (9) gives the full *36-parametric class* of the bounding media that support a particular surface plasmon mode. An important member of this class is the pair of media with no magnetoelectric coupling $\hat{X} = \hat{Y} = \hat{0}$, in which case $\hat{\epsilon} = -\hat{P}_H \hat{E}^{-1}$, $\hat{\mu} = \hat{P}_E \hat{H}^{-1}$. This means that for every novel surface plasmon mode proposed in this letter one can find a pair of simple anisotropic media, which support it.

Note, that for the sake of clarity we've been limiting our consideration by real matrices $M_{ik}^* = M_{ik}$, and real coefficients $\alpha_{ijlm}$, which ensures that that there are growing waves paired with the evanescent waves composing the surface plasmon modes and they have complex conjugated fields $\Gamma^*$. These restrictions are not essential, since $\boldsymbol{k_3}, \boldsymbol{k_4}$ and fields $\Gamma_3, \Gamma_4$ in Eq. (8) or $\boldsymbol{k_3}$ and fields $\Gamma_3$ in Eq. (9) can be selected arbitrarily. Thus consideration of complex-valued matrices $\hat{M}$ expands the *36-parametric classes* of the bounding media to *54-parametric classes* if Eq. (9) is used or to *72-parametric classes* if Eq. (8) is used.

In conclusion, we propose an algorithm to find bounding media that support surface plasmons with arbitrary frequency, wavelength, propagation direction, penetration depths and fields. Our approach shows that the any given surface plasmon mode is supported by a broad range of material pairs, which can be grouped into a *72-parametric class*.